\documentclass[%
 reprint,
 amsmath,amssymb,
 aps,
]{revtex4-2}

\usepackage{graphicx}
\usepackage{dcolumn}
\usepackage{bm}

\begin{document}

\preprint{APS/123-QED}

\title{Featuring nuanced electronic band structure in gapped multilayer graphene}

\author{Jin Jiang$^{1}$}
\author{Qixuan Gao$^{1}$}
\author{Zekang Zhou$^{1}$}
\author{Cheng Shen$^{1}$}
\author{Mario Di Luca$^{1}$}
\author{Emily Hajigeorgiou$^{1}$}
\author{Kenji Watanabe$^{2}$}
\author{Takashi Taniguchi$^{3}$}
\author{Mitali Banerjee$^{1}$}
\email{mitali.banerjee@epfl.ch}

\affiliation{$^1$ Institute of Physics, École Polytechnique Fédérale de Lausanne (EPFL), 1015 Lausanne, Switzerland}

\affiliation{
$^2$ Research Center for Electronic and Optical Materials, National Institute for Materials Science, 1-1 Namiki, Tsukuba 305-0044, Japan}

\affiliation{
$^3$ Research Center for Materials Nanoarchitectonics, National Institute for Materials Science,  1-1 Namiki, Tsukuba 305-0044, Japan}

\date{\today}

\begin{abstract}
\noindent Moiré systems featuring flat electronic bands exhibit a vast landscape of emergent exotic quantum states, making them one of the resourceful platforms in condensed matter physics in recent times. Tuning these systems via twist angle and the electric field greatly enhances our comprehension of their strongly correlated ground states. Here, we report a technique to investigate the nuanced intricacies of band structures in dual-gated multilayer graphene systems. We utilize the Landau levels of a decoupled monolayer graphene to extract the electric field-dependent bilayer graphene charge neutrality point gap. Then, we extend this method to analyze the evolution of the band gap and the flat bandwidth in twisted mono-bilayer graphene. The band gap maximizes at the same displacement field where the flat bandwidth minimizes, indicating the strongest electron-electron correlation, which is supported by directly observing the emergence of a strongly correlated phase. Moreover, we extract integer and fractional gaps to further demonstrate the strength of this method. Our technique gives a new perspective and paves the way for improving the understanding of electronic band structure in versatile flat band systems.

\end{abstract}

\maketitle


\noindent Understanding the band structure of a system is of fundamental significance in condensed matter physics. For instance, the linear conical energy spectrum of monolayer graphene (MG) gives rise to two in-equivalent K points, which results in the observation of the half-integer quantum Hall effect \cite{Novoselov,Zhang2005Nov,Zhang2006Apr}. A small twist angle between two graphene sheets induces a long-range periodic pattern, resulting in angle-dependent moiré Bloch bands \cite{Yan2012Sep}. Particularly, near a 'magic angle' ($\sim 1^\circ$) of rotation between the layers, the coupling between two graphene sheets is strongly reinforced, and the low-energy moiré bands become extremely narrow, almost without any dispersion (flat) \cite{Cao2018Apr_1}. This gives rise to exotic quantum phases, including strongly correlated insulating states \cite{Cao2018Apr_1,Liu2021Mar,Park2021Apr,Xie2019Aug,Yankowitz2019Mar,Lu2019Oct,Stepanov2020Jul,Saito2020Sep_1}, unconventional superconductivity \cite{Yankowitz2019Mar,Lu2019Oct,Stepanov2020Jul,Saito2020Sep_1,Cao2018Apr_2,Arora2020Jul,Shen2023Mar}, ferromagnetism \cite{Sharpe2019Aug,MSerlin2020Feb,Nuckolls2020Dec}, etc.

Due to the flexible carrier density control by dual electrostatic gating, twisted multilayer graphene flat band systems have become an appealing platform for studying strongly correlated quantum phases. For example, orbital Chern insulators at different integer filling factors were observed in twisted monolayer-bilayer graphene (TMBG) \cite{Chen2021Mar,Polshyn2020Dec}, while spin-polarized insulating states were observed in twisted double-bilayer graphene (TDBG) \cite{Liu2020Jul,Cao2020Jul,Shen2020May}, etc.

The dual gate tuning provides a fundamental degree of freedom to understand the rich phases mentioned above. To study these exotic phases, an experimental technique that can accurately measure the response of the device under an applied electric field is crucial. Several single-gated local probe techniques have been developed to enrich the understanding of band structure in two-dimensional materials, like scanning tunneling microscopy/spectroscopy (STM/STS) \cite{Yan2012Sep,Xie2019Aug,Shen2020May,Wong2020Jun,Jiang2019Sep,Choi2019Nov,Kerelsky2019Aug}, single electron transistor (SET) \cite{Zondiner2020Jun}, planar tunneling junction \cite{Tilak2021Jul} and nano-SQUID \cite{Uri2020May}. In addition to local probes, global measurements, including magneto-transport\cite{Cao2018Apr_1,Liu2021Mar,Park2021Apr,Xie2019Aug,Yankowitz2019Mar,Lu2019Oct,Stepanov2020Jul,Saito2020Sep_1,Cao2018Apr_2,Arora2020Jul,Shen2023Mar,Liu2020Jul,Cao2020Jul,Chen2021Mar,Polshyn2020Dec,Liu2020Jul,Cao2020Jul,Shen2020May}, nano Angle-Resolved Photoemission Spectroscopy (Nano-ARPES) \cite{Lisi2021Feb,Utama2021Feb}, electronic compressibility \cite{Tomarken2019Jul}, nano-infrared imaging \cite{Hu2017Dec} and Fourier transform infrared (FTIR) spectroscopy \cite{Ju2017Nov,Yang2022Mar} are widely adopted. However, the limiting factor in most of these aforementioned techniques, like STM/STS and nano-ARPES, is that the samples are designed only with one gate. 

\begin{figure*}
\includegraphics[width= 0.95\textwidth]{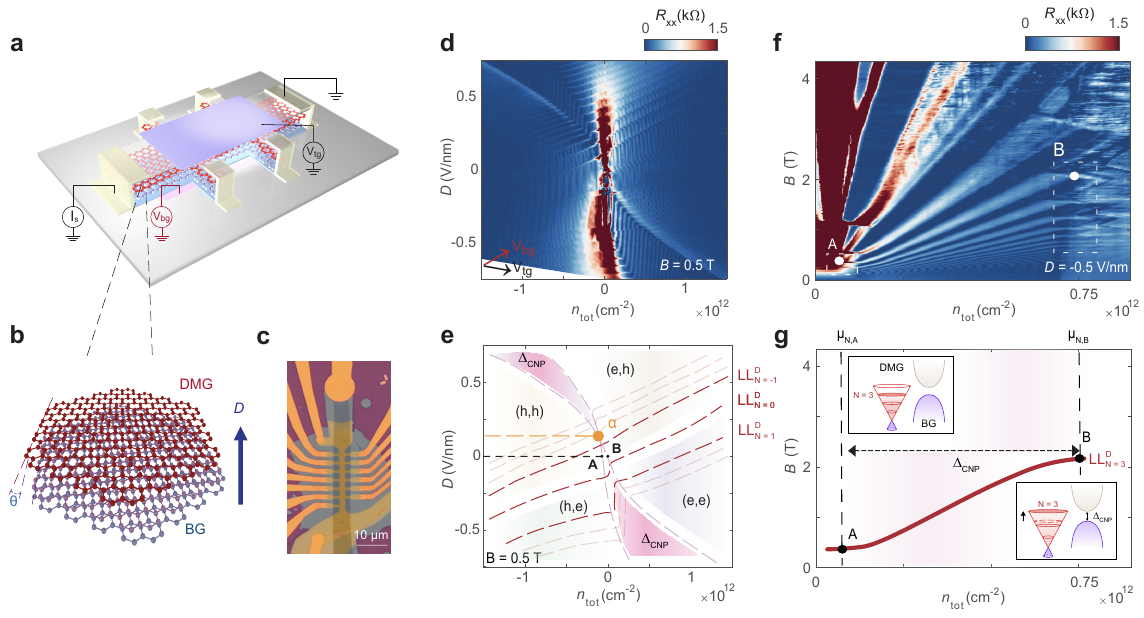}
\caption{$\textbf{Twisted decoupled monolayer graphene (DMG) with bilayer graphene (BG).}$ \textbf{a}, The schematic of the device configuration. A few nm thick graphites are used for top and bottom gates.  \textbf{b}, The schematic of DMG stacked on top of the BG. The blue arrow pointing to the DMG indicates the positive direction of the applied displacement field $D$. $\theta$ is larger than $3^\circ$ estimated from the optical image of sharp edges of DMG and BG flake. \textbf{c}, The optical image and measurement configuration of the DMG + BG device in the main text. \textbf{d}, Longitudinal resistance $R_{xx}$ as a function of the total carrier density ($n_{tot}$) and $D$ at $B = 0.5 \, \text{T}$. The red (black) arrow indicates the direction of the bottom gate (top gate). The Dirac cone of the DMG splits into many Dirac Landau levels along the top gate ($V_{tg}$) direction. \textbf{e}, A colored schematic diagram of main features of $n-D$ mapping in Fig.1d. Magenta reversed S-shaped region represents the CNP of the BG ($\Delta_{CNP}$). The two boundaries of the CNP gap of the BG are indicated by faint dashed lines intersecting at $\alpha$, where $D$ = 0.13 V/nm. $A$ and $B$ are edges of the CNP gap of the BG for the same displacement field ($D$ = 0 V/nm). A series of Dirac LLs of DMG (red dashed line along the diagonal) shuffle along the top gate direction. $LL_{N}^{D} $ is the Nth Dirac Landau level of DMG. The letters inside the parentheses represent the carrier types for BG (magenta) and DMG (blue). 'e' stands for electrons, and 'h' stands for holes. \textbf{f}, Landau fan diagram of $R_{xx}$ near the CNP of the BG at $D$ = -0.5 V/nm. White dots $A$ and $B$ represent the band crossing between the third Dirac LL of the DMG and the CNP of the BG. \textbf{g}, A schematic diagram of extracting the CNP gap of BG. The insets schematize the 3rd LL of DMG located at the edges of the CNP gap of the BG. The gap is extracted from the change in the chemical potential of LL. The chemical potential is calculated from $\mu _{N} =\mu _{N=0} + v_{F}\cdot \sqrt{2e\hbar Nsgn(N)B}$. }
\label{fig1}
\end{figure*}

Although electronic compressibility measurements,  nano-infrared imaging, Nano-SQUID-on-tip microscopy, and FTIR spectroscopy can investigate dual gate devices, these measurements each have their own drawbacks \cite{Uri2020May,Lisi2021Feb,Utama2021Feb,Hu2017Dec,Ju2017Nov,Yang2022Mar}. For example, the large beam spot ($\sim$ 1 mm) used in FTIR spectroscopy, compared with the size of typical devices, makes the experiments substantially challenging. While conventional transport measurements for extracting the energy gap, such as resistance vs. temperature ($R-T$) measurements, have been widely adopted, the thermally activated gap overshadows other details. In addition, this method cannot extract the bandwidth.

\begin{figure*}
\centering
\includegraphics[width= 0.97\textwidth]{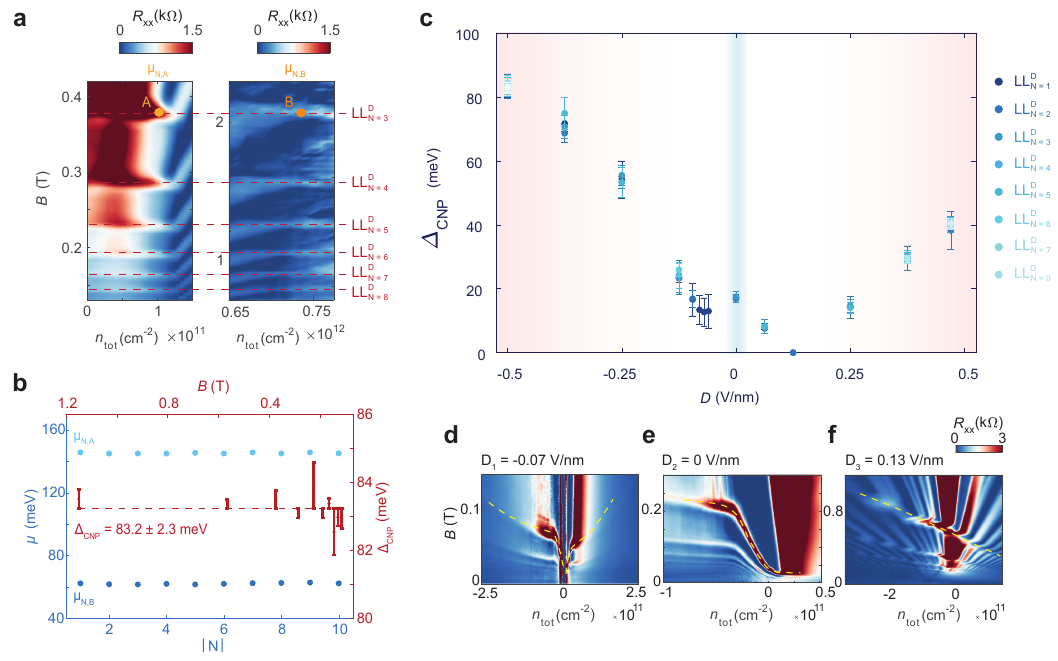}
\caption{$\textbf{Electric field-dependent energy gap extraction from DMG Landau Level spectroscopy.}$ \textbf{a}, Zooming Landau fan diagrams of the white dashed rectangular region in fig.1f. Clear Dirac LLs of the DMG indicated by red dashed lines and $LL_{N}^{D}$. The interval between LLs changes linearly with $\sqrt{B}$. \textbf{b}, The CNP gap ($\Delta_{CNP}$) of the BG extracted from different LLs of DMG. $\Delta_{CNP}$ is the change of chemical potential along the gap. The error bar is calculated from the broadening of LLs. Promoting the resolution of experiments by using the LLs with narrow bandwidth. \textbf{c}, Gap evolution as the function of $D$. The overlap of blue dots demonstrates the consistency of the gap extracted from different LLs, which increase with increasing $\left | D \right | $. The difference of Coulomb potential between two layers of BG induces a small gap at zero displacement field. \textbf{d}-\textbf{f}, Landau Level spectroscopy at different $D$. The yellow dashed line indicates the kink of the 1st LL of the DMG. The gap completely closed at a small positive electric field $D_3$. }
\label{fig2}
\end{figure*}

In this work, we report a simple technique to study the electronic band structure of electric field-tunable systems. To achieve this, we utilize the Landau levels (LLs) of monolayer graphene as a sensor (Landau level spectroscopy) to extract the electric field-tunable bilayer graphene (BG) charge neutrality point (CNP) gap. We measured a gap of 83.2 ± 2.3 meV at $D$ = -0.5 V/nm , and the accuracy of this result surpasses the precision of previously reported conventional nano-ARPES measurements\cite{Lisi2021Feb,Utama2021Feb}. Furthermore, to demonstrate the versatility of this technique, we have also measured the bandwidth and the band gap as a function of the external displacement field for the TMBG flat band system. We are also able to extract integer and fractional flat band gaps without being influenced by the thermal activation energy, as compared to conventional $R-T$ measurements.

\section{Extracting the electric field-tunable CNP gap in bilayer graphene}

\noindent To benchmark our technique, we performed electrical transport measurements on a dual-gated Hall bar fabricated on a monolayer graphene(MG) stacked on a bilayer graphene (BG) with a large twist angle, as shown in Fig. 1a, 1b. The large twist angle $\theta$ between MG and BG is necessary to decouple the two layers. The electrical transport measurement and corresponding colored schematic diagram are shown in Fig. 1d and 1e, respectively. 

We trace the CNP of this decoupled MG (DMG). This is denoted by the red dashed line in Fig. 1e, $LL_{N= 0}^{D}$, and N is the Landau level index. In our device configuration, the BG is closer to the bottom gate. Hence, the bottom gate cannot tune the DMG due to the screening from the BG, which is evident from the trace of the CNP line of the DMG as it is almost parallel to the bottom gate direction. Moreover, we observe multiple Landau levels splitting under a finite magnetic field (see $n-D$ at zero magnetic field diagram of more devices in Fig.S1). These Dirac Landau levels of the DMG ($LL_{N}^{D}$) shuffle along the $V_{tg}$ direction when $V_{bg}$ is fixed. This asymmetric gate response is referred to as "layer-specific anomalous screening (LSAS)" \cite{Uri2023Aug} and is an intrinsic screening behavior resulting from our device geometry. For a conventional BG device, the CNP peak should track $n_{tot}$ = 0 and follow the hBN thickness ratio ($d_{b}$/$d_{t}$, where $d_{b}$, $d_{t}$ are the thicknesses of the bottom and the top hBN respectively). Here, we can see that $\Delta_{CNP}$ (reversed S-shaped magenta region in Fig.1e) deviates from the $n_{tot}$ = 0. 

\begin{figure*}
\centering
\includegraphics[width= 0.95\textwidth]{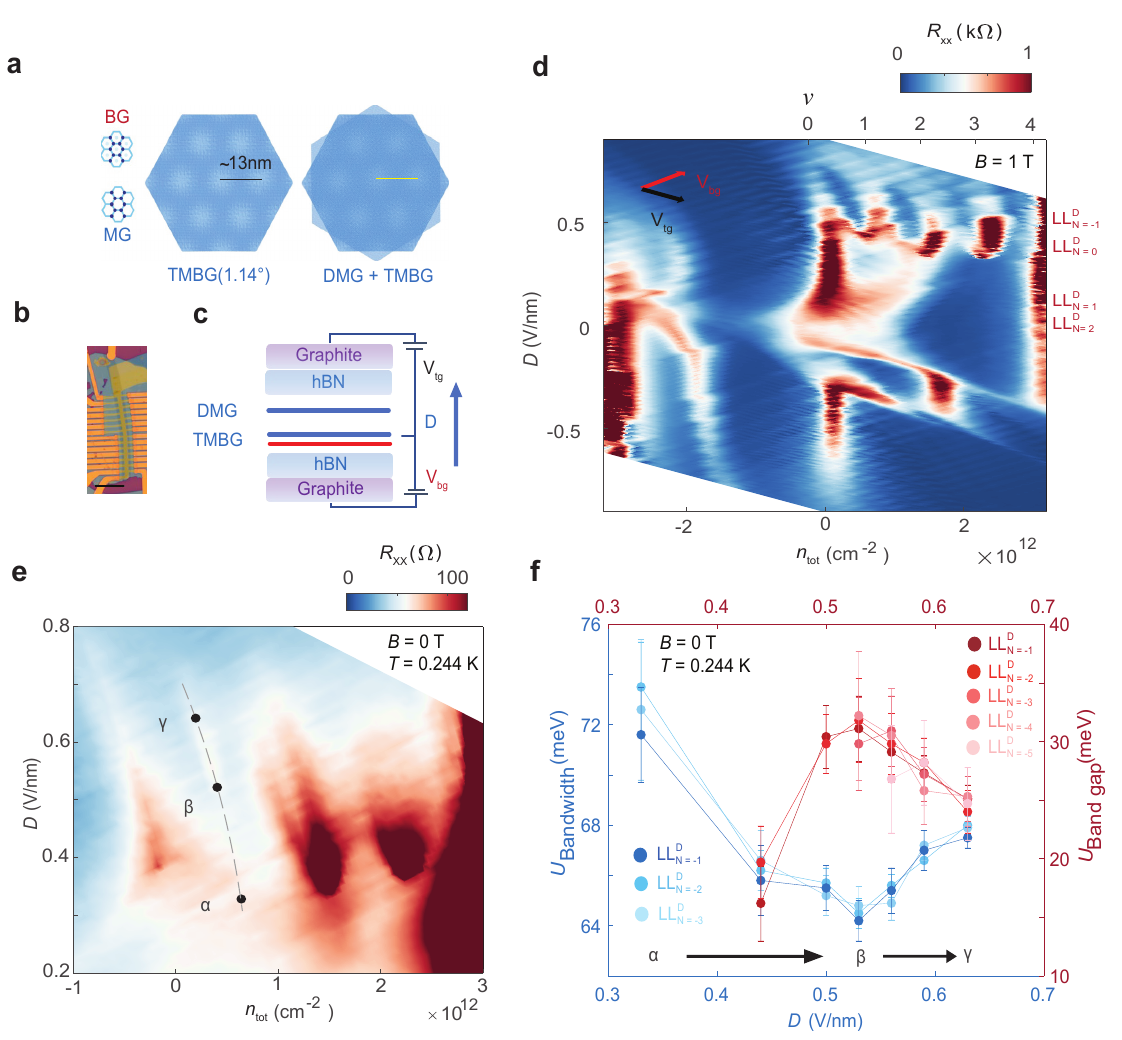}
\caption{$\textbf{Featuring the strongest correlation in twisted mono-bilayer graphene}$. \textbf{a}, Schematics of TMBG and DMG with TMBG. The left panel is a moiré pattern of magic angle TMBG (1.14° $\pm$ 0.02°). The right panel is DMG on top of the TMBG, indicating the same moiré length as TMBG. \textbf{b}, Optical image of device. The black scale bar is 10 $\mu$m. \textbf{c}, The configuration of dual gate measurements. The blue arrow pointing to the DMG indicates the positive direction of the applied displacement field $D$. \textbf{d}, $R_{xx}$ as a function of the $n_{tot}$ and $D$ at $B = 1 \, \text{T}$. The asymmetric phase dependent on $D$ is similar to a phase of pure TMBG, indicating that the monolayer graphene is fully decoupled from BG. The correlated states appear at all integer filling factors at positive $D$. \textbf{e}, Zooming of the $n-D$ map presented in Fig.4d, at $B = 0 \, \text{T}$. \textbf{f}, Fingerprint of correlations strength in relation to TMBG flat bandwidth and band gap at different $D$. The minimum of the bandwidth and the maximum of the band gap at $D$ = 0.53 V/nm ($\beta$) indicate the strongest electron-electron correlations, where exactly a $\nu=1$ correlated state emerges in Fig. 3e. }
\label{fig3}
\end{figure*}

Using Landau level spectroscopy, we estimated the $\Delta_{CNP}$ value. Since the DMG is decoupled from the BG, we can directly estimate the chemical potential extracted from the $LL_{N}^{D}$, which depends precisely on the applied magnetic field. Fig. 1f shows the band crossings between the $LL_{N}^{D}$ and the CNP gap edges of the BG denoted by $A$ and $B$. In order to estimate $\Delta_{CNP}$, we take $LL_{N=3}^{D}$ (indicated by the red solid lines in Fig.1g) as an example. $\mu _{N, A}$, $\mu _{N, B}$ denote chemical potentials of the left and right CNP edges (indicated by two blue dashed lines and inset schematics), following $\mu _{N} =\mu _{N=0} + v_{F}\cdot \sqrt{2e\hbar N sgn(N)B}$, where Fermi velocity $v_{F} = 1\times 10^{6} m/s$ and $\hbar$ is reduced Planck constant. The chemical potential difference $\Delta_{N} = \mid \rm \mu _{N, A} - \mu _{N, B}\mid \rm$ represents the energy gap size of the bilayer graphene CNP.

In Fig.2a, at both points $A$ and $B$, the chemical potential extracted through the different $LL_{N}^{D}$ changes as ${\sqrt N}$, implying that $A$ and $B$ are both tracking the same DMG Landau levels, indicated by red dash lines. Fig.2b illustrates the measured $\Delta_{CNP}$. For different $LL_{N}^{D}$, $\Delta_{CNP}$ values remain consistent, with small variation. Furthermore, due to the narrow width of the $LL_{N}^{D}$ shown in Fig.2a, the experimental error is smaller ($< \text{few meV}$) than the previously reported values, showcasing the high precision of this method. We could increase the experimental resolution by a finer magnetic field scanning.

We then extend our analysis to ianclude various displacement fields ($D$), allowing us to observe the electric-field dependence of $\Delta_{CNP}$ in BG, as shown in Fig. 2c. The $\Delta_{CNP}$ in BG increases significantly with increasing  $\left | D  \right | $  as a result of the difference in Coulomb potential between the layers (See details in Fig.S2). The estimated $\Delta_{CNP}$ at  $D$ = -0.5 V/nm is 83.2 ± 2.3 meV, is in agreement with previous findings \cite{Zhang2009Jun, Iwasaki2022Oct}.
\begin{figure}
\centering
\includegraphics[width= 0.45\textwidth]{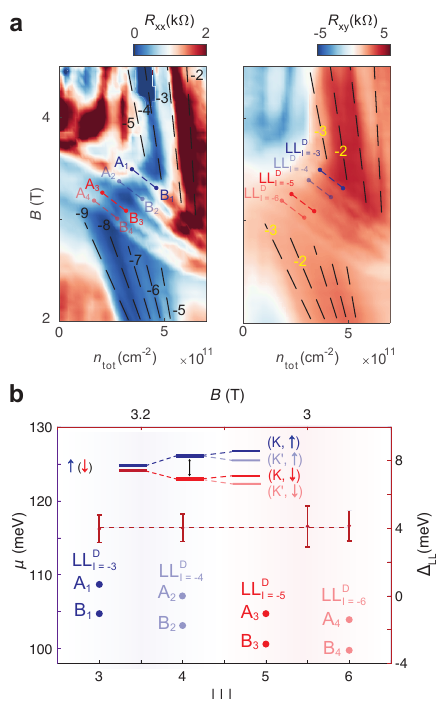}
\caption{$\textbf{Flat band integer Landau Level gap}$. \textbf{a}, Longitudinal resistance $R_{xx}$ and Hall resistance $R_{xy}$ as a function of $n_{tot}$ and $B$. The Dirac LLs degeneracy is lifted, and $LL_{N = -1}^{D}$ splits into four mini-bands $LL_{l}^{D}$ indicated by the purple and red dashed lines. The black dashed lines indicate the flat band LLs. As DMG LLs offer Chern number offset, the corresponding Chern number $C$ (indicated by black number) equals $ C_{F}$ (indicated by yellow number) + $C_{D}$, $C_{F}$ and $C_{D}$ are Chern numbers of TMBG flat band and DMG Dirac band respectively. \textbf{b}, The flat band integer Landau level gap ($LL_{l= -2}^{f}$). Multiple $A$s, $B$s represent gap edges of $LL_{l= -2}^{f}$ and the estimated gap $\bigtriangleup_N$ = $\mid \mu _{A} - \mu _{B} \mid$\rm  = 4.1 $\pm $ 0.9 meV is averaged through different $LL_{l}^{D}$.
}
\label{fig4}
\end{figure}

We observed a finite gap (17.2 ± 1.3 meV) at the CNP at $D$ = 0 V/nm, indicating the inversion symmetry is spontaneously broken in BG, which results from layer polarization of the valence and conduction bands \cite{McCann2006Oct,McCann2007Jul,Zhu2022Nov}. This $\Delta_{CNP}$ is in good agreement with the theoretical work in a twisted multilayer graphene system where the gap size saturates to $16$ meV as a result of a large twist angle ($ > 3^{\circ} $) \cite{Zhu2022Nov}. Furthermore, we observed the evolution of the CNP gap closing and reopening with increasing positive $D$ (See details in Fig.S3). Particularly, at $D_3$ = 0.13 V/nm, the gap is completely closed by sign change of Coulomb potential difference \cite{Ohta2006Aug}, which is indicated by the  $\alpha$ point in Fig.1e. This behavior results from the asymmetric response of the displacement field and layer polarization. The weaker response of the CNP gap to positive $D$ (pointing to MG, $D$ = 0.25 V/nm, $\Delta_{CNP}$ is 14.3 ± 2.7 meV) compared to negative $D$ (pointing to BG, $D$ = -0.25 V/nm, $\Delta_{CNP}$ is 55.3 ± 4.0 meV) supports this observation. In addition, we observe that the CNP gap does not fully close within the moderate negative $D$ range ($D_2$ to $D_1$). Further investigation is needed into the band structure, particularly considering the overlap between the CNP of the DMG and the BG. (See details in Fig.S4).

\section{Characterizing the strongest correlation in twisted mono-bilayer graphene}

In order to establish the robustness of this technique, we study a more complicated electric field-tunable flat band system-TMBG. The DMG was placed on top of the monolayer graphene of the magic angle TMBG ($1.14^{\circ }  \pm 0.02^{\circ }$) as shown in Fig. 3a. The schematic compares the moiré patterns of TMBG and DMG+TMBG, the moiré length is equal to $\sim 13$ nm and does not change significantly with the addition of the DMG. A $n-D$ map at $B$ = 1 T is shown in Fig. 3d (See data for more devices in Fig.S5). We observe similar multiple Landau levels $LL_{N}^{D}$ splitting as in Fig. 1d. Then the strongly correlated states are observed at all integer filling factors and are found to vary with the $D$. For TMBG, the spatial inversion symmetry is broken; it exhibits a rich phase diagram similar to the one of twisted bilayer graphene (TBG) when a $D$ is applied in the direction of MG (TBG side). The TMBG exhibits a phase diagram similar to that of twisted double bilayer graphene (TDBG) when the $D$ is inverted \cite{Chen2021Mar,Polshyn2020Dec}.

\begin{figure*}
\centering
\includegraphics[width= 0.94\textwidth]{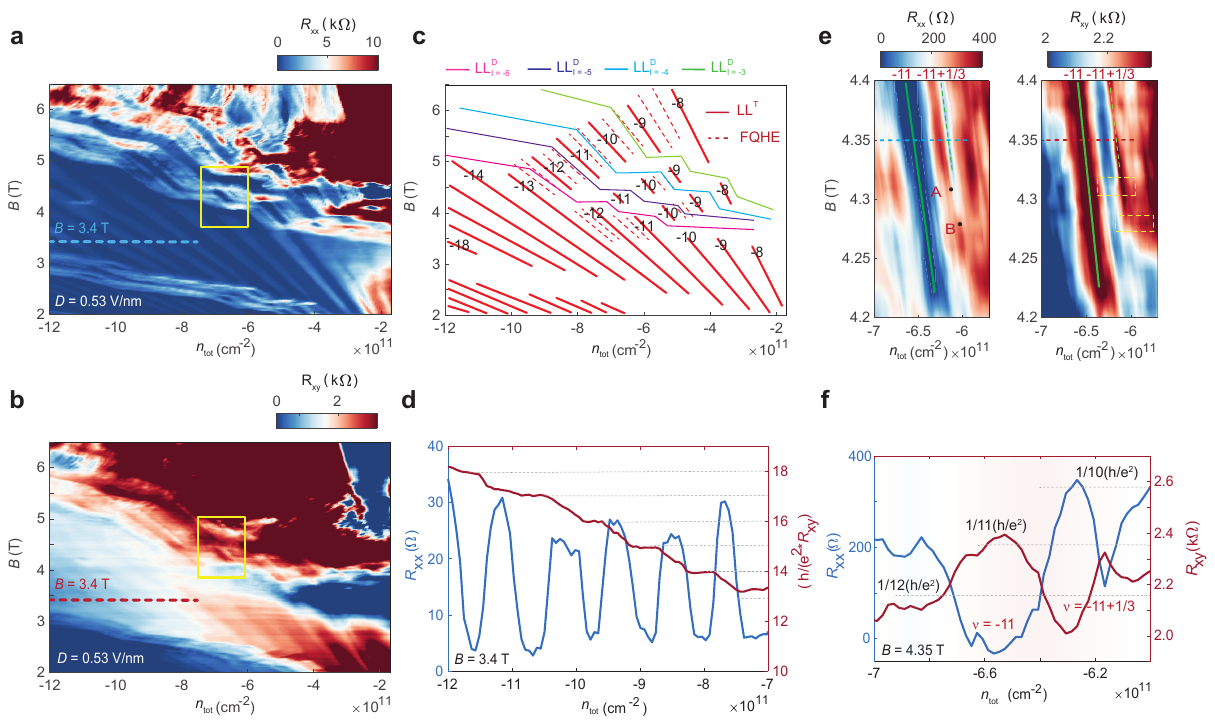}
\caption{$\textbf{Chern number cascade and a fractional gap.}$ \textbf{a},\textbf{b}, Longitudinal resistance $R_{xx}$ and Hall resistance $R_{xy}$ as a function of $n_{tot}$ and $B$. \textbf{c}, Schematic of quantum oscillation of Fig. 5a. $LL_{}^{T}$ is TMBG flat band LLs composing with DMG LLs Chern number offset. \textbf {d}, Filling factor dependence of longitudinal resistance (blue dash line in Fig.5a) and Hall resistance plateaus (red dash line in Fig.5b). \textbf{e}, zoomed Landau fan taken from yellow rectangular in Fig. 5a, and Fig. 5b. The solid green line indicates $\nu = |-11| $ integer quantum Hall state.  The green dashed line denotes a fractional quantum Hall state of $\nu = |-11 + 1/3 |$. The black dash line indicates the band crossing between the fractional flat band $\nu_{f} = |-6 + 1/3 |$ and the Dirac band $LL_{l= -5}^{D}$ ($C_{D}=-5$). $A$, $B$ are band edges of the fractional gap of $\nu_{f} = |-6 + 1/3 |$. The estimated gap $\bigtriangleup_N$ = $\mid \mu _{A} - \mu _{B} \mid$\rm  = 0.56 $\pm $ 0.14 meV. \textbf{f}, Evidence of a fractional quantum Hall state. The data of $R_{xx}$ and $R_{xy}$ are taken along the blue and red dash lines in Fig. 5e.
}
\label{fig5}
\end{figure*}

Figure 3e is a zoom-in $n-D$ map (TBG-side) at $B$ = 0 T. Figure 3f shows the extracted flat bandwidth and band gap as a function of $D$ (See details of this analysis in Extended Fig.1). When the correlated state appears at point $\alpha$ ($D$ = 0.33 V/nm) with a filling factor of 2, 3, the flat bandwidth is estimated to be about 73.5 $\pm$ 1.8 meV, which is in good agreement with the experimental work in TMBG (70 $\pm$ 10 meV) \cite{Zhang2024May}. From point $\alpha$ to point $\beta$ ($D$ = 0.53 V/nm), the flat bandwidth decreases to 64.5 $\pm$ 0.6 meV as $D$ increases, which shows that the $D$ greatly suppresses the width of the flat band. Simultaneously, the band gap varies inversely with the $D$. Particularly, at point $\beta$, the flat bandwidth reaches a minimum while the band gap maximizes, and a correlated state appears at filling factor $\nu = 1$, as shown in Fig. 3e, indicating the strongest electron-electron correlation. From $\beta$  to $\gamma$, the flat bandwidth broadens with increasing $D$ and decreasing band gap, reflecting weaker electronic correlations. According to the previous works \cite{Chen2021Mar,Polshyn2020Dec}, the flat band gradually touches the remote dispersive band as the displacement field increases. (See more details in Fig.S6 to Fig.S9)

\section{Flat band integer and  fractional quantum Hall gaps}

\noindent In order to further warrant the accuracy of this technique, we demonstrate how this method can be used to extract integer and fractional quantum Hall gaps for the TMBG flat band. (In Fig.S10, we show the high homogeneity of our device). The four-fold degeneracy of LLs for both DMG and the flat band is lifted. In Fig. 4a, we clearly see four mini-bands $LL_{-l}^{D}$ formed from $LL_{N = -1}^{D}$ splitting, indicated by red and purple dashed lines. These DMG Landau levels with different filling factors induce additional Chern number offsets (-l) (See for more detailed analysis Figs. S11, S12, and S13). The total Chern number $C$ (indicated by the black numbers) is the summation of $ C_{f}$ (indicated by the yellow numbers) and $C_{D}$, where $C_{f}$ and $C_{D}$ are Chern numbers of TMBG flat band and DMG Dirac band respectively. Multiple $A$s and $B$s represent gap edges of $LL_{l= -2}^{f}$ and the estimated gap $\bigtriangleup_N$ = $\mid \mu _{A} - \mu _{B} \mid$ is averaged through different $LL_{l}^{D}$ in Fig. 4b. The extracted values of the gap measured by different LLs are consistent with each other, and the average value of the gap is given by 4.1 ± 0.9 meV. 

This integer LL gap at $\nu = |-2 |$ increases with increasing magnetic field. The estimated gap size is  $ 1.17$ $\pm $ 0.26 meV/T to  $ 1.28$ $\pm $ 0.23 meV/T. These results are comparable with reported works on BG of a gap at $\nu = |-2|$ ($1.2$ meV/T)\cite{Martin2010Dec}. Although the magnetic field dependence of the LLs gap differs between BG and TMBG, further investigation is required to determine the evolution of the flat band gap under different $B$ and $D$ using different DMG LLs.

The schematic inside Fig.4b shows a larger separation between the $LL_{l= -4}^{D}$ and $LL_{l= -5}^{D}$, suggesting that the energy gap between two spin states ($\downarrow /\uparrow$) is greater than the gap between different valleys (K/K'). The priority for filling spin states over valley states arises from the Zeeman effect \cite{Zhang2006Apr}.

We extend a similar analysis to a fractional gap. Figures 5a and 5b show $R_{xx}$ and $R_{xy}$ as a function of $B$ and $n_{tot}$ near the CNP region. The schematic of the corresponding Landau level crossings is shown in Fig. 5c. The N = -1 Dirac Landau level splits into four mini bands, indicated by the green (Landau level filling factor l = -3), blue (l = -4), purple (l = -5), and pink (l = -6) lines, respectively. Thus, the Chern number cascade was observed when crossed with the flat-band Landau levels. Figure 5d exhibits near zero $R_{xx}$ (blue line) and corresponding well-quantized $R_{xy}$ plateaus (red line); these are line cuts taken along the blue and red dash line in Fig. 5a, 5b. A difference of neighboring $LL_{N}^{D}$ filling factor is one, indicating that the degeneracy of the system is fully lifted.

According to the band edges $A$ and $B$ shown in Fig. 5e, a fractional gap can be extracted from the band crossing between the fractional flat band $\nu_{f} = |-6 + 1/3 |$ and the Dirac band $LL_{l= -5}^{D}$ ($C_{D}=-5$). The magnetic fields in Fig. 5e at $A$ and $B$ are 4.310 $\pm $ 0.005 T and 4.270 $\pm $ 0.005 T, respectively.
 Thus, the extracted flat band fractional gap is $\bigtriangleup_N$ = $\mid \mu _{A} - \mu _{B} \mid$\rm  = 0.56 $\pm $ 0.14 meV. At filling factors $\nu = \mid -11\mid$ and $\nu = \mid -11 + 1/3\mid$, $R_{xx}$ exhibits a minimum (blue line) and $R_{xy}$ (red line) approaches the quantized plateaus $-11(e^2/h)$ as shown in Fig. 5f. This guides us to define the edges of the FQHE band in Fig.5e. Our fractional gap with $ 0.13$ $\pm $ 0.03 meV/T at $D$ = 0.53 V/nm is comparable with reported works on BG of a gap at $\nu = |1/3 |$ ($0.12$ meV/T to $ 0.15$ meV/T )\cite{Feldman2012Sep} and a gap at $\nu = |2/3 |$ ($0.16$ meV/T) \cite{Shi2016Feb}.

\section{Conclusion}

\noindent  To summarize, we have used the Landau levels of a decoupled monolayer graphene to measure the chemical potential of dual-gated multilayer graphene devices. First, we measured the electric field-tunable CNP gap of bilayer graphene. Then, we extracted the electric field-tunable flat bandwidth and band gap in twisted mono-bilayer graphene. This resourceful technique enables us to find a connection between the flat bandwidth and the strongest correlation, which is supported by the direct observation of a strongly correlated phase emerging. Moreover, the measurements of the flat band integer and fractional gaps provide a promising avenue to investigate nuanced band structure.

This technique has far-reaching consequences for studying strongly correlated states. For example, the superconducting phase diagram and the ground state can be understood by studying adjacent correlated states (See more details in Fig.S14). Currently, there is a scarcity of techniques that establish a connection between the displacement field-tunable flat bandwidth and strong electron-electron correlations. Our work can encourage more theoretical works to understand the complicated phase diagrams of these multilayer graphene systems. Furthermore, it could be extended in the future to other similar moiré systems, such as transition metal dichalcogenides systems.

\nocite{*}

\bibliography{apssamp}

\section*{METHODS}

\textbf{\begin{center}Device Fabrication\end{center}}

\noindent The devices are fabricated using an advanced technique known as "cut and stack" \cite{Saito2020Sep_2}. First, pristine materials such as monolayer graphene, bilayer graphene, hBN (10-50 nm), and graphite flakes (3-15 nm) were mechanically exfoliated on an oxygen plasma-etched $\rm SiO_{2}$ (285 nm thick) surface. Next, we used Atomic Force Microscopes(AFM) to pre-cut monolayer graphene and bilayer graphene. High-quality homogeneous poly (bisphenol A carbonate) (PC)/polydimethylsiloxane (PDMS) was then stacked on the glass slide used to transfer the 2D materials flakes to the alignment marker chip. The transfer stage precisely controls the twisted angle between two 2D materials to within 0.1° resolution. The graphite top gate is then fabricated, followed by the electrodes by electron beam lithography and metal evaporation. Here, we use the conventional etching method to define Hall bars. We etch graphite and hbN with $\rm O_{2}$ and $\rm SF_{6}$ gases, respectively. Optimization of the etching parameters is important to obtain 1D edge contacts with the Cr/Au (5/50 nm) electrodes \cite{Wang2013Nov}.

\textbf{\begin{center}Measurements\end{center}}

\noindent Transport measurements were performed in cryostat Heliux with a base temperature of around 240 mK. Standard locking techniques were implemented using the Stanford Research SR860 with an excitation frequency of f = 17.7777 Hz and an AC excitation current of less than 10 nA. The transport measurements are performed in a four-terminal geometry.We can extract $n_{tot}$ and the displacement field $D$ using the following equation $n_{tot} = V_{bg}C_{bg}/e + V_{tg}C_{tg}/e$,  $\rm D = \mid V_{bg}C_{bg} - V_{tg}C_{tg}\mid/(2\varepsilon _{0} )$. $C_{bg}$, $C_{tg}$ are the capacitances between the top ($V_{tg}$) and bottom ($V_{bg}$) gates and the DMG + TMBG, e is the electron charge and $\varepsilon _{0}$ is the vacuum permittivity.

\textbf{\begin{center}Twisted angle determination\end{center}}

\noindent Due to the decoupling of the DMG from the TMBG, the carrier density $n_{flat-band}$ observed by $\nu = \pm 4 = \pm n_{s}$ of band insulating states. This value can be determined by analyzing quantum oscillations in the Landau fan diagram or by making observations in an $n-D$ diagram. n as function of twisted angle following equation $n_{s}= 8\theta ^{2} /\sqrt{3} a^{2}$  determines twisted angle of our TMBG(1.14° $\pm$ 0.02°), a = 0.246 nm is lattice constant of graphene. The same rule applies to the determination of small-angle TMBG, as shown in Fig.S5 and Fig.S14.

\

\begin{acknowledgments}

\noindent We thank Oleg Yazyev, Yifei Guan, and Thomas Ihn for important discussions. J.J. acknowledges funding from SNSF. M.B. acknowledges the support of SNSF Eccellenza grant No. PCEGP2\_194528, and support from the QuantERA II Programme that has received funding from the European Union’s Horizon 2020 research and innovation program under Grant Agreement No 101017733. K.W. and T.T. acknowledge support from the JSPS KAKENHI (Grant Numbers 20H00354 and 23H02052) and World Premier International Research Center Initiative (WPI), MEXT, Japan.

\end{acknowledgments}

\textbf{\begin{center}Author Contribution\end{center}}

\noindent J.J. and M.B. conceived the project. J.J. fabricated the devices. Q.G. fabricated one device in the supplementary file. J.J. performed the measurements with the help of Z.Z.. J.J. has analysed the data with inputs from C.S. and Z.Z.. K.W. and T.T. provided the hBN crystals. J.J. wrote the manuscript with inputs from all authors.

\textbf{\begin{center}Competing Interests\end{center}}

\noindent The authors declare that they have no competing interests.

\textbf{\begin{center}Data availability\end{center}}

\noindent All data are available from the corresponding authors on reasonable request.

\setcounter{figure}{0}
\renewcommand{\figurename}{Extended Fig}
\renewcommand{\tablename}{Extended Data Table}

\begin{figure*}
\centering
\includegraphics[width= 1\textwidth]{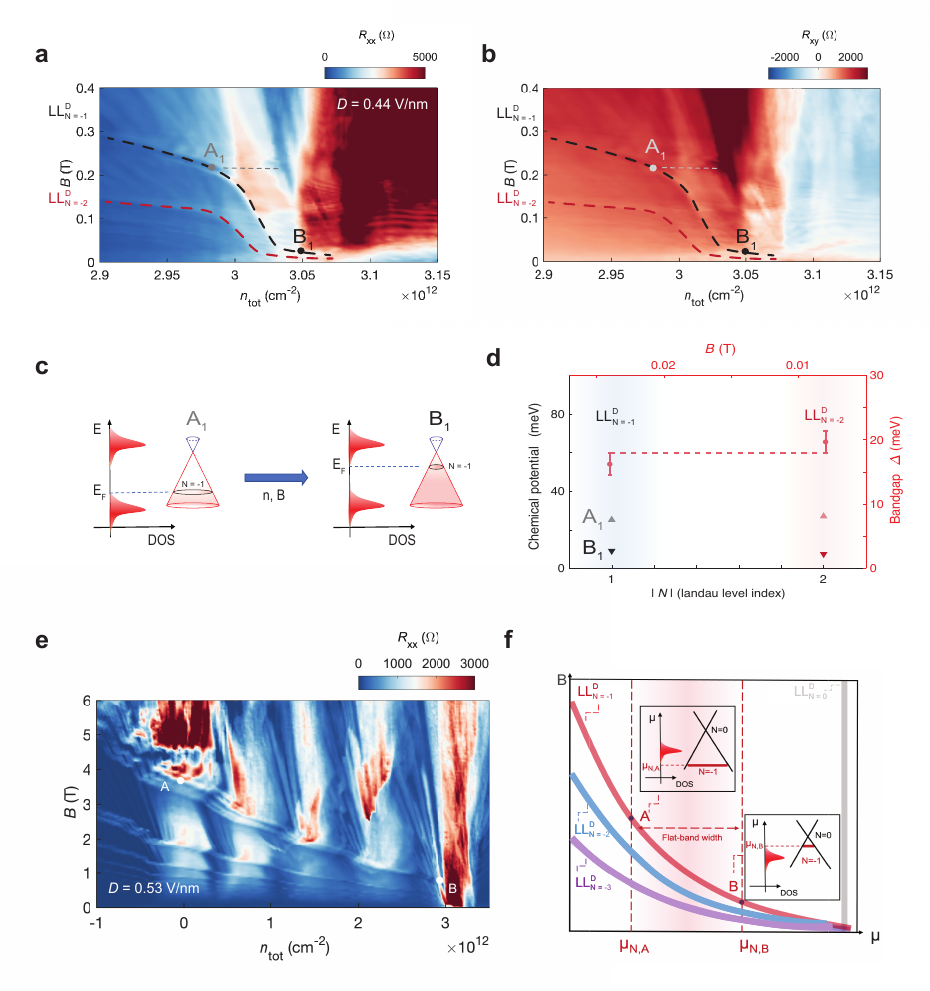}
\caption{\textbf{ Band gap and Bandwidth extraction of twisted DMG and TMBG device.} \textbf{a}, \textbf{b}, Landau fan diagrams near full filling factor 4 (band gap). Black and red dash line indicate $LL_{N = -1}^{D}$, $LL_{N = -2}^{D}$, respectively. $A_{1} (B_{1})$ represent the chemical potential on the bottom (top) side of the band gap. \textbf{c}, Schematic diagram of extracting flat band gap. $\Delta_{N} = \mid \rm  \mu _{N,A_{1}} - \mu _{N,B_{1}}\mid \rm$ represents the band gap of the TMBG. \textbf{d}, The band gap is determined by the $LL_{N = -1}^{D}$, $LL_{N = -2}^{D}$,. Since the band structure hardly varies with the magnetic field, the difference between the band gaps derived from two different Dirac LLs is very small, which means that our method is reliable. \textbf{e}, Landau fan diagram of $R_{xx}$ up to full filling with D = 0.53 V/nm. $A$ and $B$ represent the edges of the flat band in TMBG. \textbf{f}, Schematic diagram of extracting flat bandwidth from $LL_{N}^{D}$. The $A$ ($B$) represents the chemical potential on the bottom (top) side of the bandwidth. $\Delta_{N} = \mid \rm \mu _{N,A} - \mu _{N,B}\mid \rm$ represents the bandwidth of the TMBG.}
\label{Extended Fig.3}
\end{figure*}


%
%
%
%
%
%
\setcounter{figure}{0}
\renewcommand{\figurename}{Fig}
\renewcommand{\tablename}{Extended Data Table}

\onecolumngrid

\newpage

\clearpage

\section*{Supplementary Information}

\noindent In Fig.S1, we made two twisted decoupled monolayer graphene (DMG) bilayer graphene (BG) devices; we consistently observe similar phase diagrams in transport measurements. For example, we observe that the $LL_{N = 0}^{D}$ (DMG charge neutrality point (CNP)) is nearly parallel to the bottom gate ($V_{bg}$) direction as shown in the yellow dashline in Fig.S1a and Fig.S1d. In the magnetic field, the Dirac cone splits into many Dirac Landau levels $LL_{N}^{D}$ along the top gate ($V_{tg}$) direction. The $V_{bg}$ is anomalously weakened when tuning the DMG as it is screened by the BG, and the same principle applies to the weakening of $V_{tg}$ when tuning the BG.

\

In Fig.S5, we made four twisted DMG and TMBG (0.46°, 0.71°, 1.14° and 1.64°) devices. In all four devices, whether the bottom TMBG is small twist angle (SATMBG), magic angle (MATMBG) or large twist angle TMBG (LATMBG); we consistently observe that the $LL_{N = 0}^{D}$ is nearly parallel to the bottom gate direction as shown in Fig.S1.  

\

In Fig.S11 and Fig.S12, starting from a-b, at 0.5 T, a Dirac Landau level of $LL_{N = 0}^{D}$ appears, indicated by the black dashed line. Next, in Fig.S12, when the magnetic field reaches 2 T, this N = 0 Landau level breaks down into two two-fold isospin polarization Landau levels. Then, at 3 T, the two two-fold bands successively breaks into 4 finer Landau levels $LL_{l}^{D}$, similar result shown in Fig.S11.

\

In Fig.S13, we can observe that it is not only the 4-fold degeneracy of the Dirac cone is lifted, but also the 4-fold degeneracy of the flat bands at all integer fillings 0, 1, 2, 3, and 4 is lifted. The strong Coulomb interaction can lift the degeneracy of four-fold isospin flavors, such as spin and valley (orbital), leading to nontrivial band topology by generating spontaneous symmetry breaking at different integer fillings.

\

In Fig.S13a, 13c, we find that the Landau level filling factors of DMG are negative, indicating that the DMG Dirac cone is in a hole-doped state. This is consistent with our previous observation. Moreover, in Fig.S13a, the Landau fan can be divided into different regions at the edges of the Dirac cone Landau levels, denoted by I, II, and III. At all integer filling factors (including CNP, band insulator), the flat-band Landau level ($LL_{N}^{}$) acquires additional Chern numbers ranging from 2, 6, 10 to 4N + 2, corresponding to I, II and III to N region, respectively. The 4N + 2 Chern number offset is because the Dirac cone in the hole-doped state provides an additional Chern number of -(4N + 2). A similar Chern number shift effect is observed on the TDBG side of our device, as shown in Fig.S13b, 13d. So the total Chern number $C$ equals $ C_{f}$ + $C_{D}$, $C_{f}$ and $C_{D}$ are Chern numbers of TMBG flat band and DMG Dirac band respectively.

\

In Fig.S14, we made three twisted DMG and TBG (0.94°, 1.01° and 1.20°) devices. The existence of decoupled monolayer
graphene at one side of the target multilayer graphene systems will inevitably introduce the asymmetry which can significantly affect the electronic band structure . The layer polarization generated from the carrier density difference between different layers of the target multilayer graphene systems will lead to gap opening or band flattening. From this point of view, our technique has far-reaching consequences beyond merely measuring bandwidth and gaps. We can study how the layer polarization and screening
affect on strongly correlations and superconductivity in twisted bilayer graphene,which arise from tuning the flat band. 

\begin{figure*}
\centering
\renewcommand{\thefigure}{S1}
\includegraphics[width= 1\textwidth]{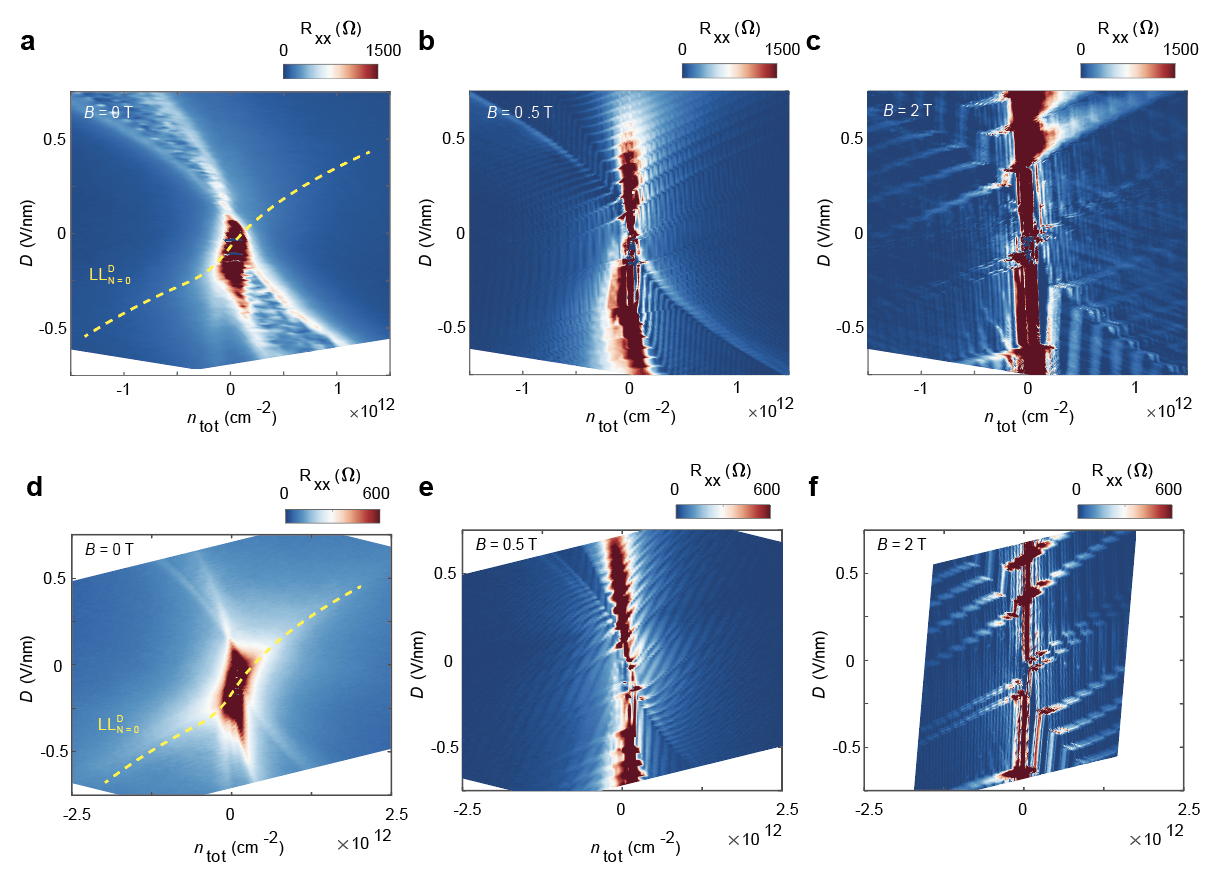}
\caption{$\textbf{Twisted decoupled monolayer graphene (DMG) bilayer graphene (BG) devices.}$ \textbf{a}-\textbf{c}, Longitudinal resistance $R_{xx}$ as a function of the total carrier density ($n_{tot}$) and displacement fields ($D$) of Device A at $B = 0 \, \text{T}$, $B = 0.5 \, \text{T}$ and $B = 2 \, \text{T}$, respectively). The yellow dashed line indicates the CNP of the DMG. \textbf{d}-\textbf{f}, Same Transport measurements of Device B.}
\label{figS1}
\end{figure*}

\begin{figure*}
\centering
\renewcommand{\thefigure}{S2}
\includegraphics[width= 0.6\textwidth]{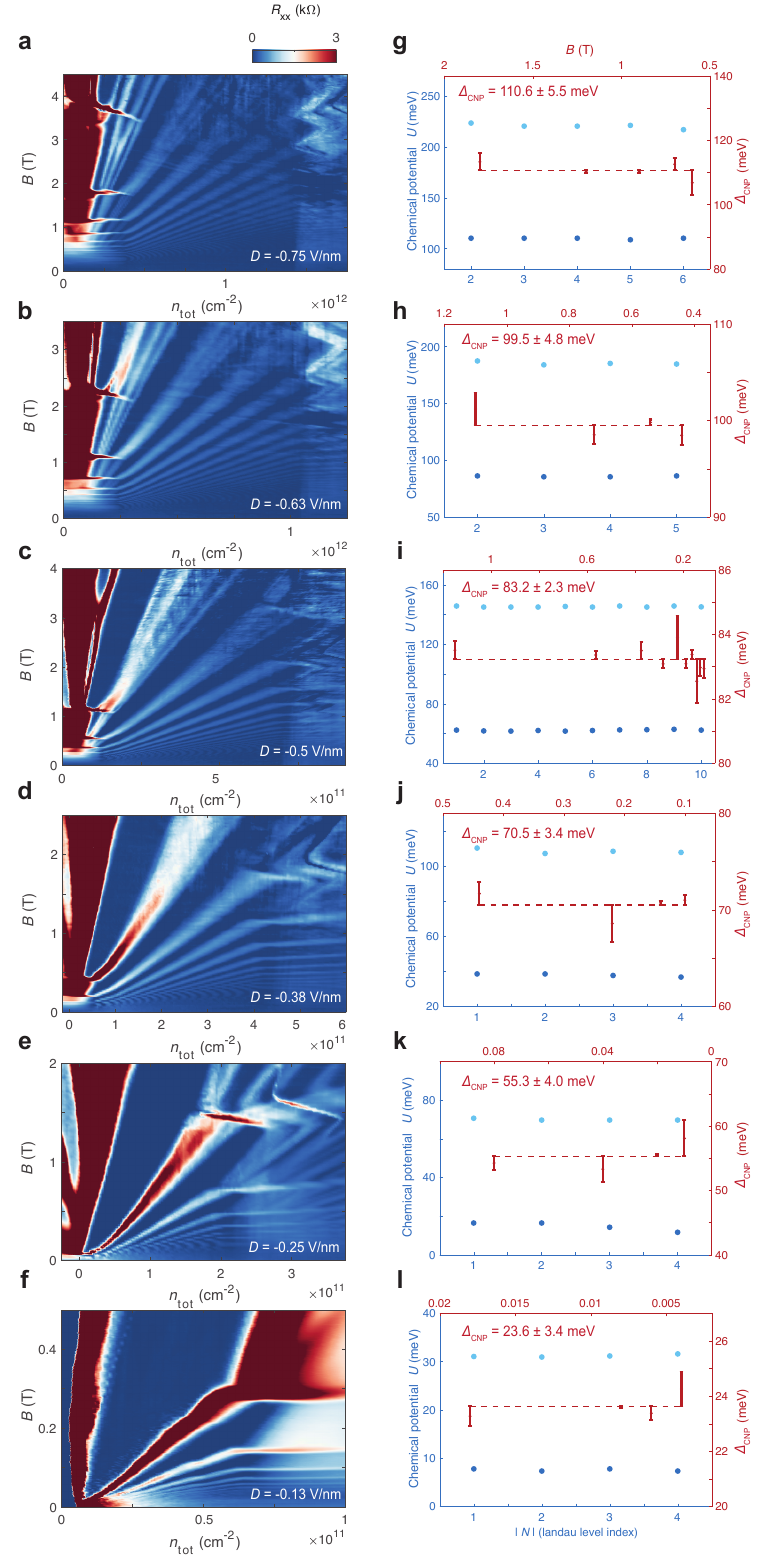}
\caption{$\textbf{The evolution of BG CNP gap under various negative displacement fields.}$ \textbf{a}-\textbf{f}, Quantum oscillations at various negative displacement fields (pointing to BG). \textbf{g}-\textbf{l}, The corresponding $\Delta_{CNP}$ calibrated with different $\rm D_{N}$. The $\Delta_{CNP}$ in BG increases significantly with increasing $D$.
}
\label{figS2}
\end{figure*}

\begin{figure*}
\centering
\renewcommand{\thefigure}{S3}
\includegraphics[width= 0.7\textwidth]{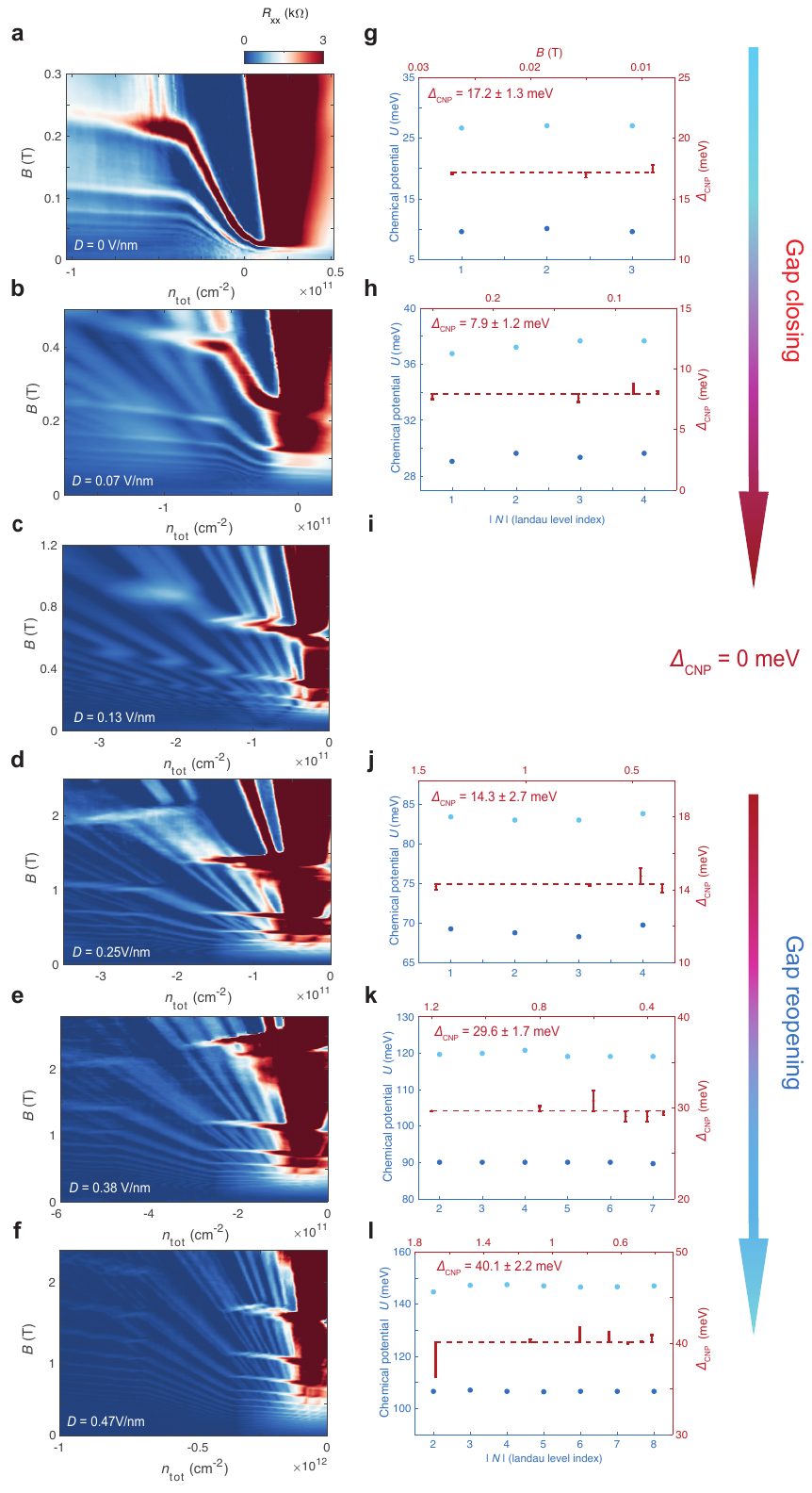}
\caption{$\textbf{The BG CNP gap closing and reopening.}$ \textbf{a}-\textbf{c}, Quantum oscillations at zero and various positive displacement fields (pointing to DMG). \textbf{g}-\textbf{l}, The corresponding $\Delta_{CNP}$ calibrated with different $\rm D_{N}$. The $\Delta_{CNP}$ in BG is closing and reopening. At $D$ = 0.13 V/nm, the gap is completely closed. This is due to the sign change in the Coulomb potential difference, which results from the asymmetric response of layer polarization within the system.
}
\label{figS3}
\end{figure*}

\begin{figure*}
\centering
\renewcommand{\thefigure}{S4}
\includegraphics[width= 1\textwidth]{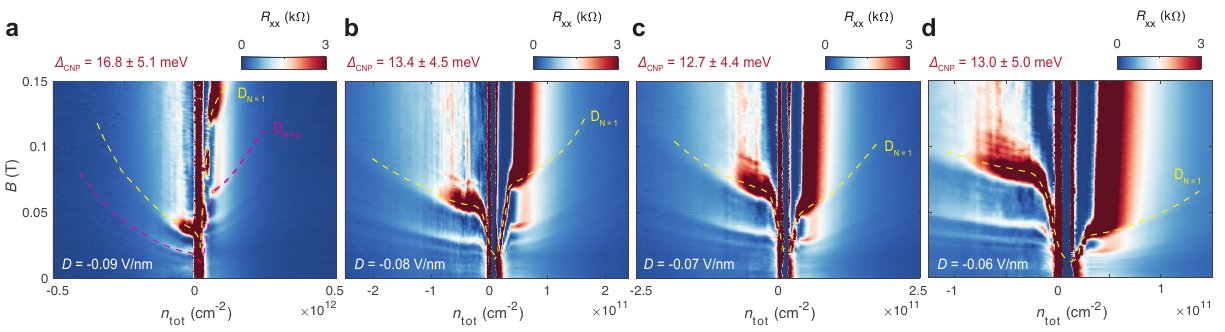}
\caption{$\textbf{The evolution of BG CNP gap under moderate negative displacement fields.}$ \textbf{a}-\textbf{d}, Quantum oscillations at moderate negative displacement fields (pointing to BG). The yellow dashed line indicates the kink of the 1st LL of the DMG. The DMG CNP moves when the displacement fields of the system are tuned.
}
\label{figS4}
\end{figure*}

\begin{figure*}
\centering
\renewcommand{\thefigure}{S5}
\includegraphics[width= 1\textwidth]{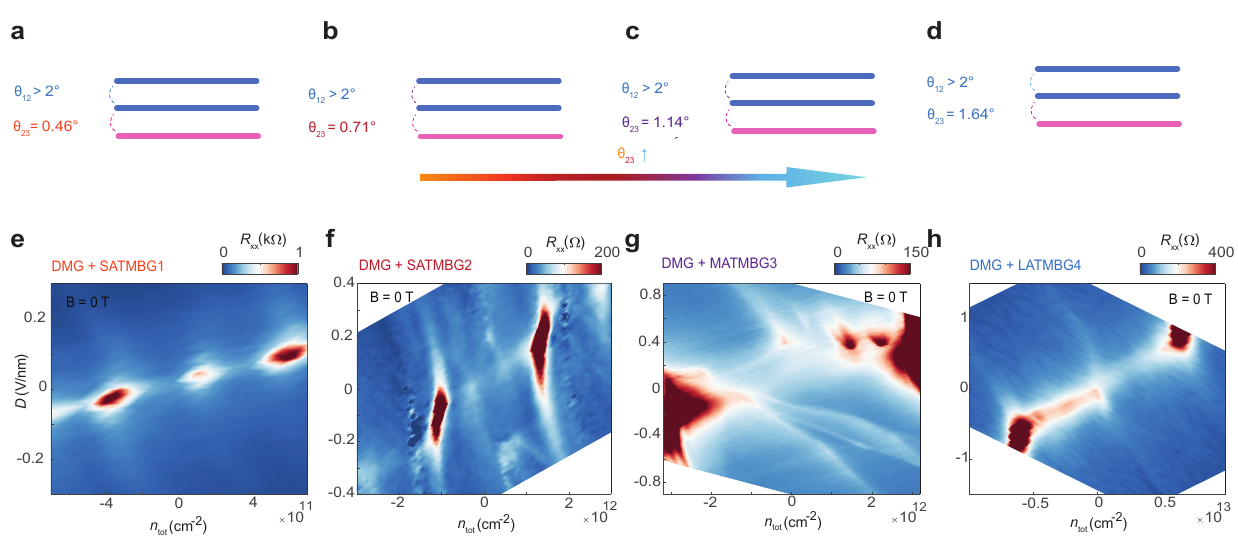}
\caption{$\textbf{Twisted DMG and twisted monolayer bilayer graphene (TMBG) devices.}$ \textbf{a}-\textbf{d}, Schematic diagrams of four DMG stacked on twisted monolayer bilayer graphene (TMBG) devices (0.46°, 0.71°, 1.14° and 1.64°)  at $B = 0 \, \text{T}$, respectively. \textbf{e}-\textbf{h},  $n-D$ maps of devices corresponding to a-d).
}
\label{figS5}
\end{figure*}

\begin{figure*}
\centering
\renewcommand{\thefigure}{S6}
\includegraphics[width= 0.8\textwidth]{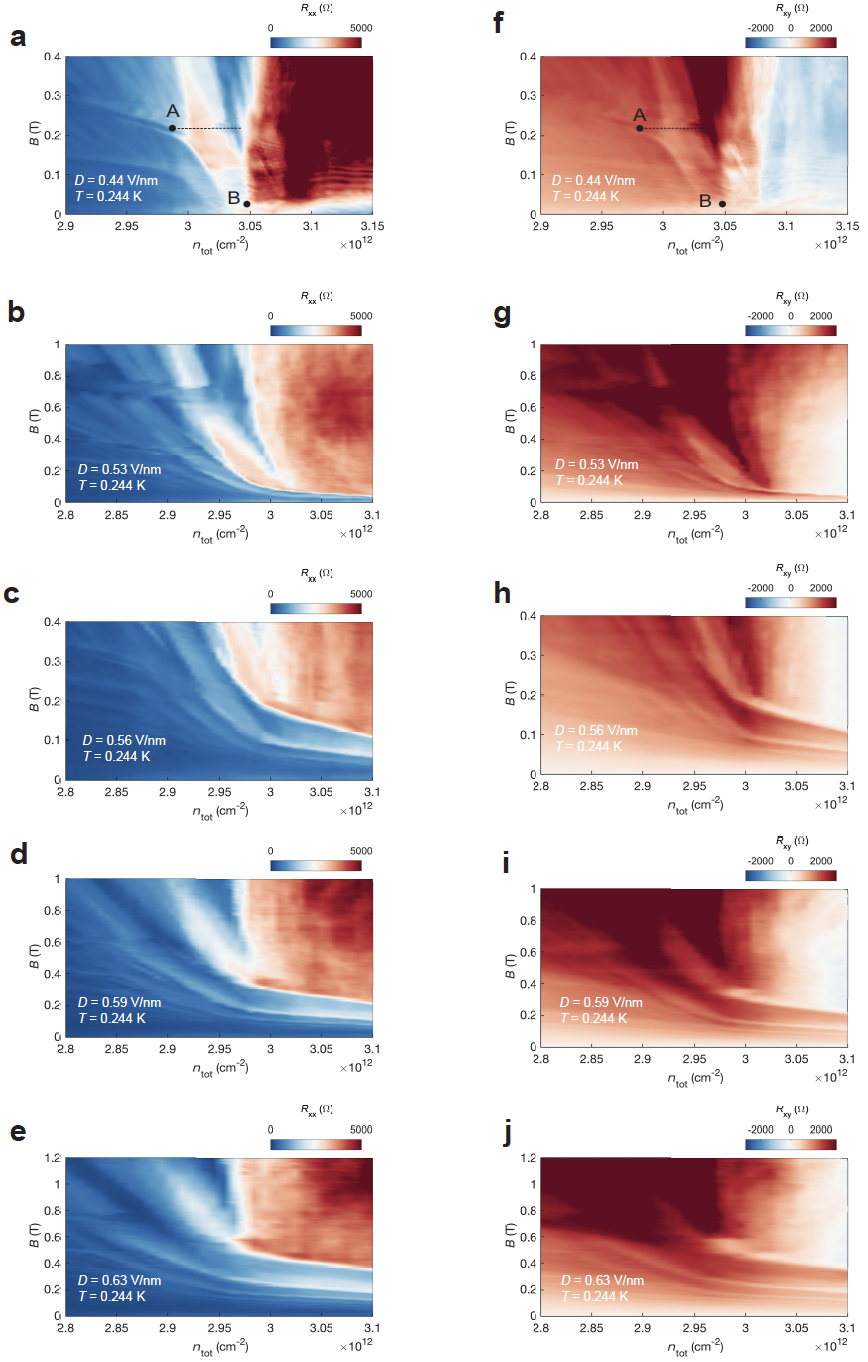}
\caption{$\textbf{Landau fan diagrams near band gap under different displacements fields.}$ \textbf{a}-\textbf{e}, Landau fan diagrams near full filling factor 4 (band gap) of a decoupled monolayer graphene (DMG) stacked on twisted monolayer bilayer graphene (1.14° $\pm$ 0.02°) device at different displacement fields. A  and B denote the band edges of the bandgap. \textbf{f}-\textbf{j},  Same as a-e, only Hall resistances are different.
}
\label{figS6}
\end{figure*}

\begin{figure*}
\centering
\renewcommand{\thefigure}{S7}
\includegraphics[width= 0.8\textwidth]{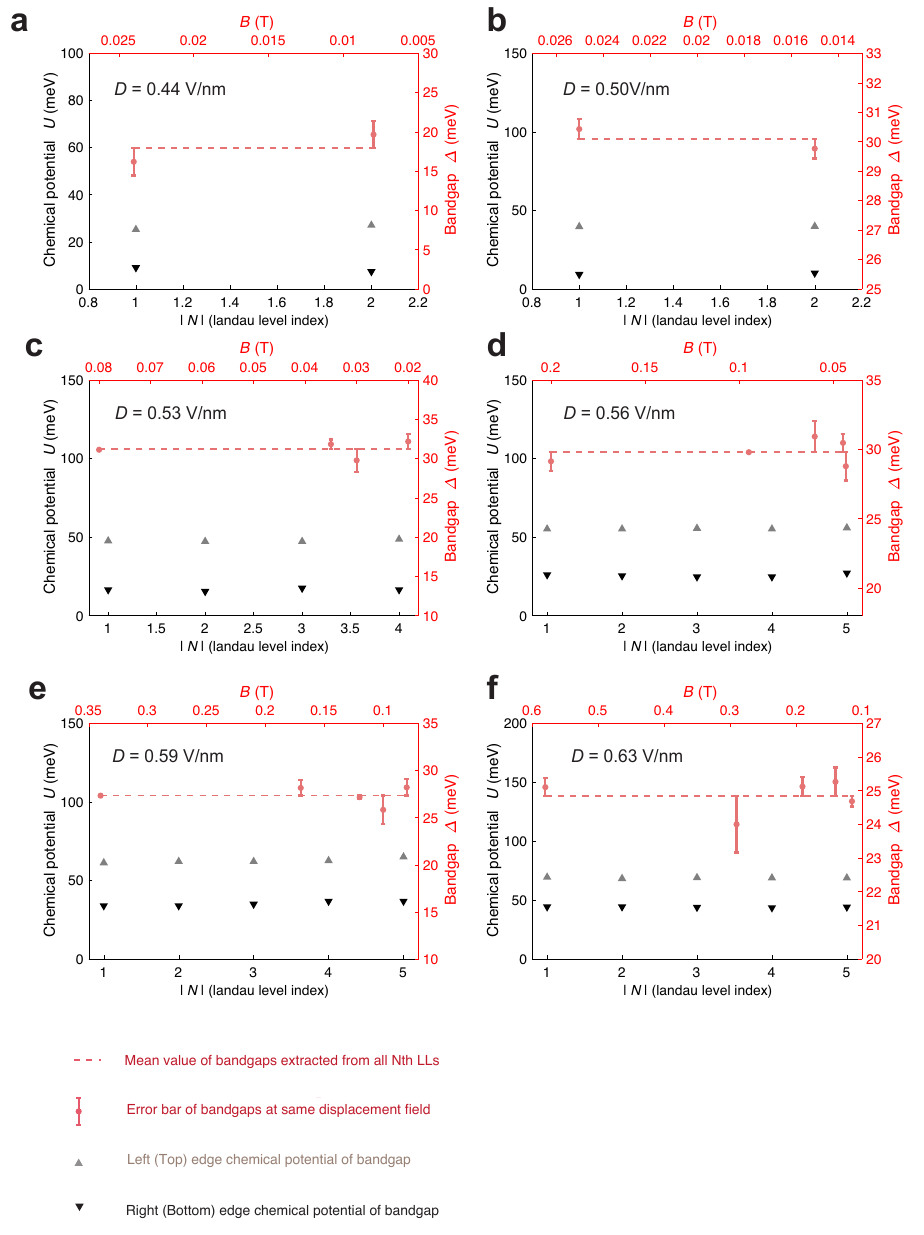}
\caption{$\textbf{Band gap extraction under different displacements fields.}$ \textbf{a}-\textbf{f}, The bandgap is determined by the different $\rm D_{N}$ at different displacement fields. From $D$ = 0.44 V/nm to $D$ = 0.63 V/nm.
}
\label{figS7}
\end{figure*}

\begin{figure*}
\centering
\renewcommand{\thefigure}{S8}
\includegraphics[width= 1\textwidth]{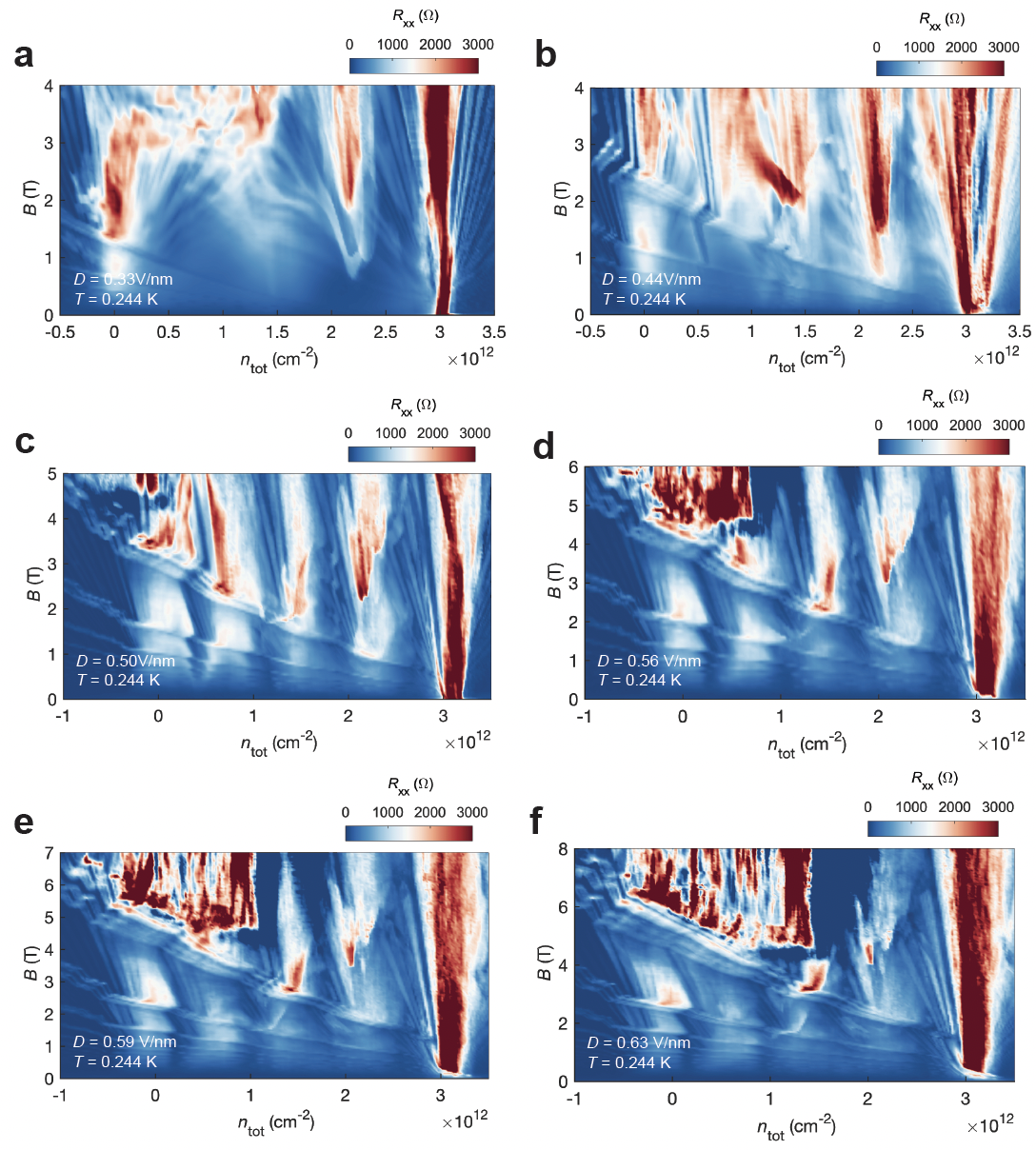}
\caption{$\textbf{Landau fan diagrams of TMBG flat band under different displacements fields.}$ \textbf{a}-\textbf{f}, Landau fan diagrams of a decoupled monolayer graphene (DMG) stacked on twisted monolayer bilayer graphene (1.14° $\pm$ 0.02°) device at different displacement fields.
}
\label{figS8}
\end{figure*}

\begin{figure*}
\centering
\renewcommand{\thefigure}{S9}
\includegraphics[width= 0.8\textwidth]{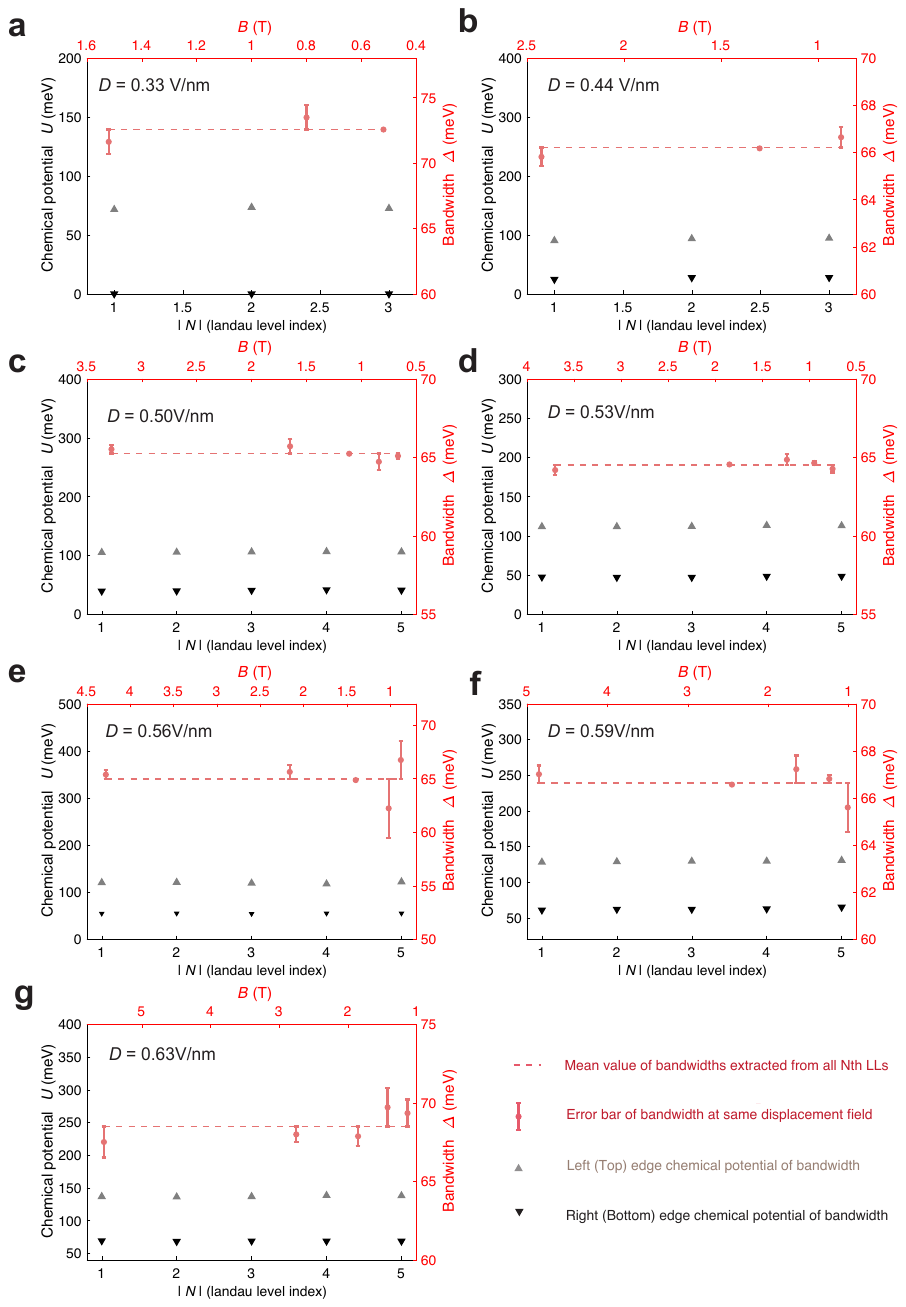}
\caption{$\textbf{TMBG flat bandwidth extraction under different displacements fields.}$\textbf{a}-\textbf{g}, The bandwidth is determined by the different $\rm D_{N}$ at different displacement fields. From $D$ = 0.33 V/nm to $D$ = 0.63 V/nm.
}
\label{figS9}
\end{figure*}

\begin{figure*}
\centering
\renewcommand{\thefigure}{S10}
\includegraphics[width= 0.8\textwidth]{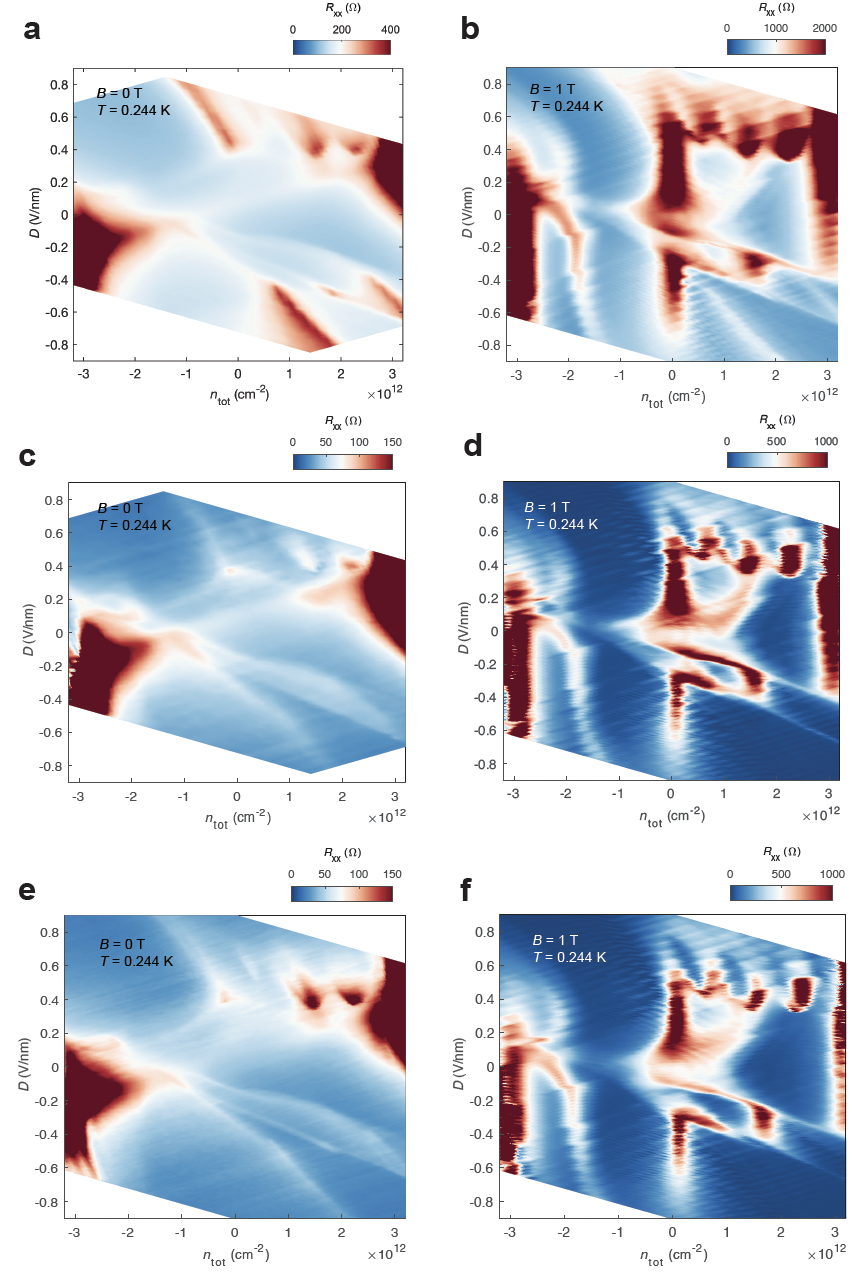}
\caption{$\textbf{High homogeneity of device.}$ \textbf{a}-\textbf{c}, $n-D$ maps for different contact pairs of a decoupled monolayer graphene (DMG) stacked on twisted monolayer bilayer graphene (1.14° $\pm$ 0.02°) device at $B = 0 \, \text{T}$. \textbf{d}-\textbf{f}, Same as a-c, only $B = 1 \, \text{T}$ is different. Transport results at different contact points show the high homogeneity of this device.
}
\label{figS10}
\end{figure*}

\begin{figure*}
\centering
\renewcommand{\thefigure}{S11}
\includegraphics[width= 1\textwidth]{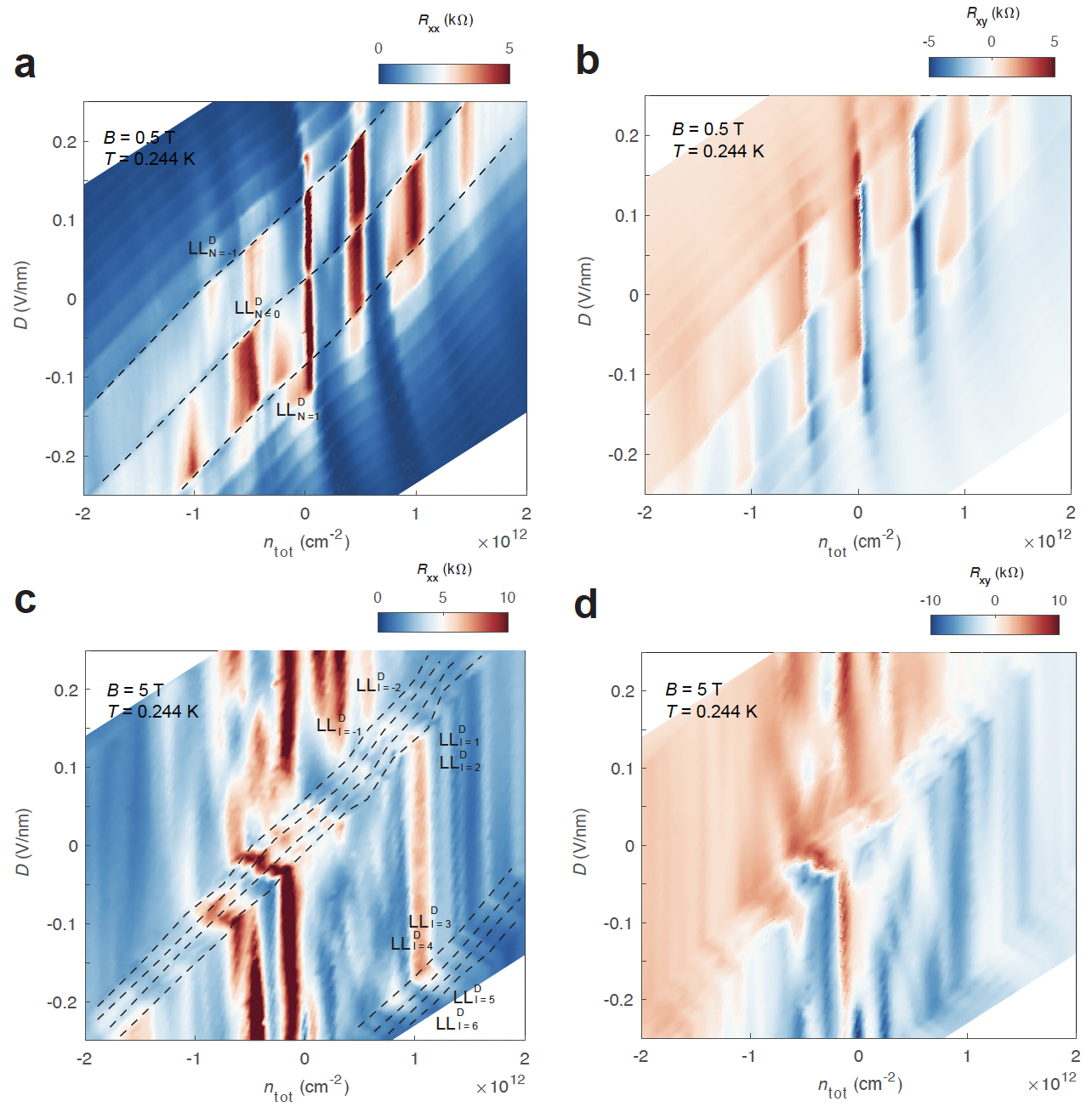}
\caption{$\textbf{Dirac Landau levels degeneracy lifting of twisted DMG SATMBG device.}$ \textbf{a},\textbf{b}, Phase diagram of a decoupled monolayer graphene (DMG) stacked on twisted monolayer bilayer graphene (0.46° $\pm$ 0.05°) device at $B = 0.5 \, \text{T}$. \textbf{c},\textbf{d}, Same as a-b, only $B = 5 \, \text{T}$ is different.
}
\label{figS11}
\end{figure*}

\begin{figure*}
\centering
\renewcommand{\thefigure}{S12}
\includegraphics[width= 0.95\textwidth]{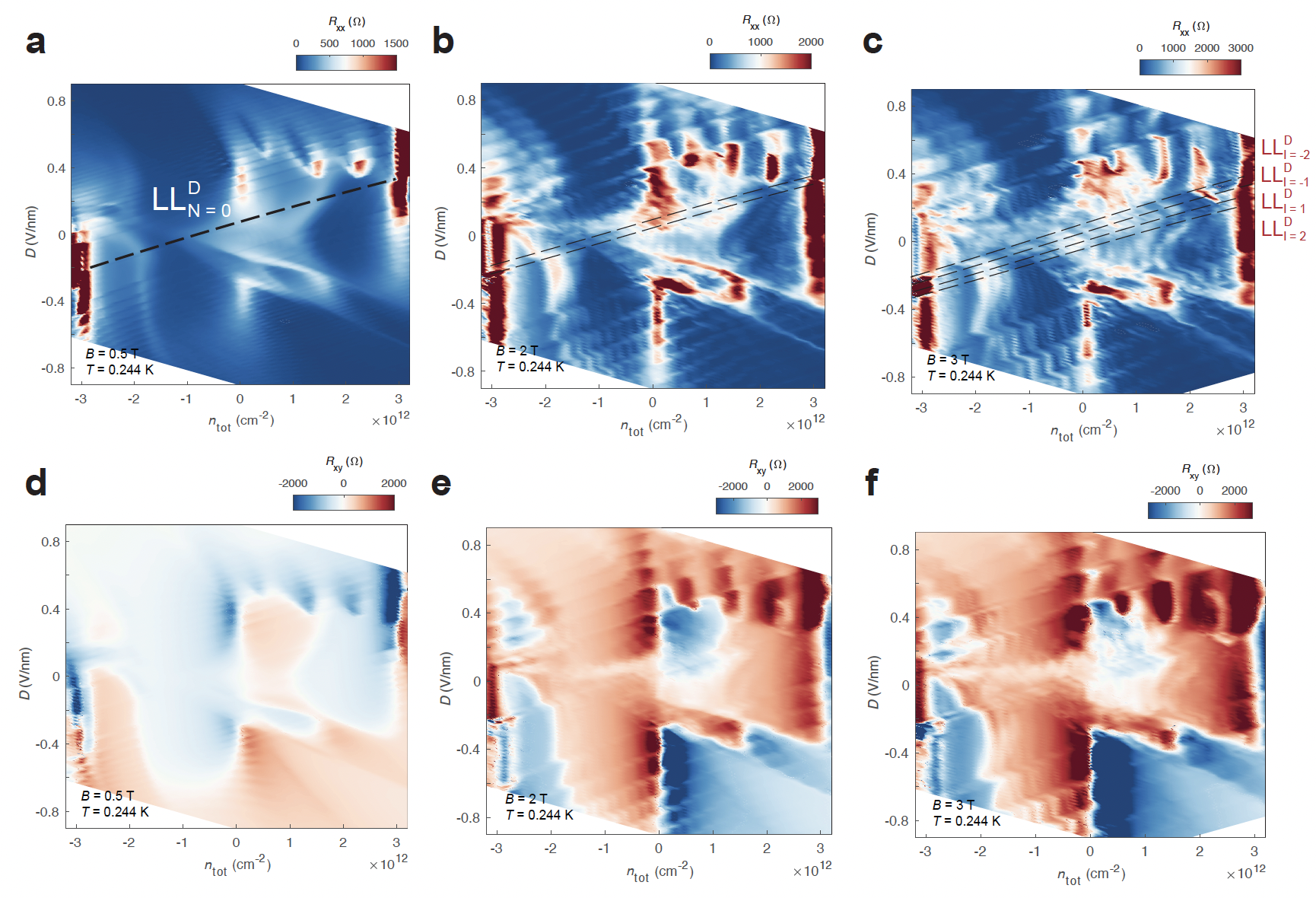}
\caption{$\textbf{Dirac Landau levels degeneracy lifting of twisted DMG MATMBG device.}$ \textbf{a}-\textbf{c}, Phase diagrams of a decoupled monolayer graphene stacked on twisted monolayer bilayer graphene (1.14° $\pm$ 0.02°) device at different magnetic fields, $B$ = 0.5 T, $B$ = 2 T, $B$ = 3 T. The phase diagrams are expressed in terms of longitudinal and Hall resistances, which are a function of the total carrier density, $n_{tot}$, and the vertical displacement field, D, respectively. \textbf{d}-\textbf {c}, Same as a-b with measuring Hall resistance.}
\label{figS12}
\end{figure*}

\begin{figure*}
\centering
\renewcommand{\thefigure}{S13}
\includegraphics[width= 1\textwidth]{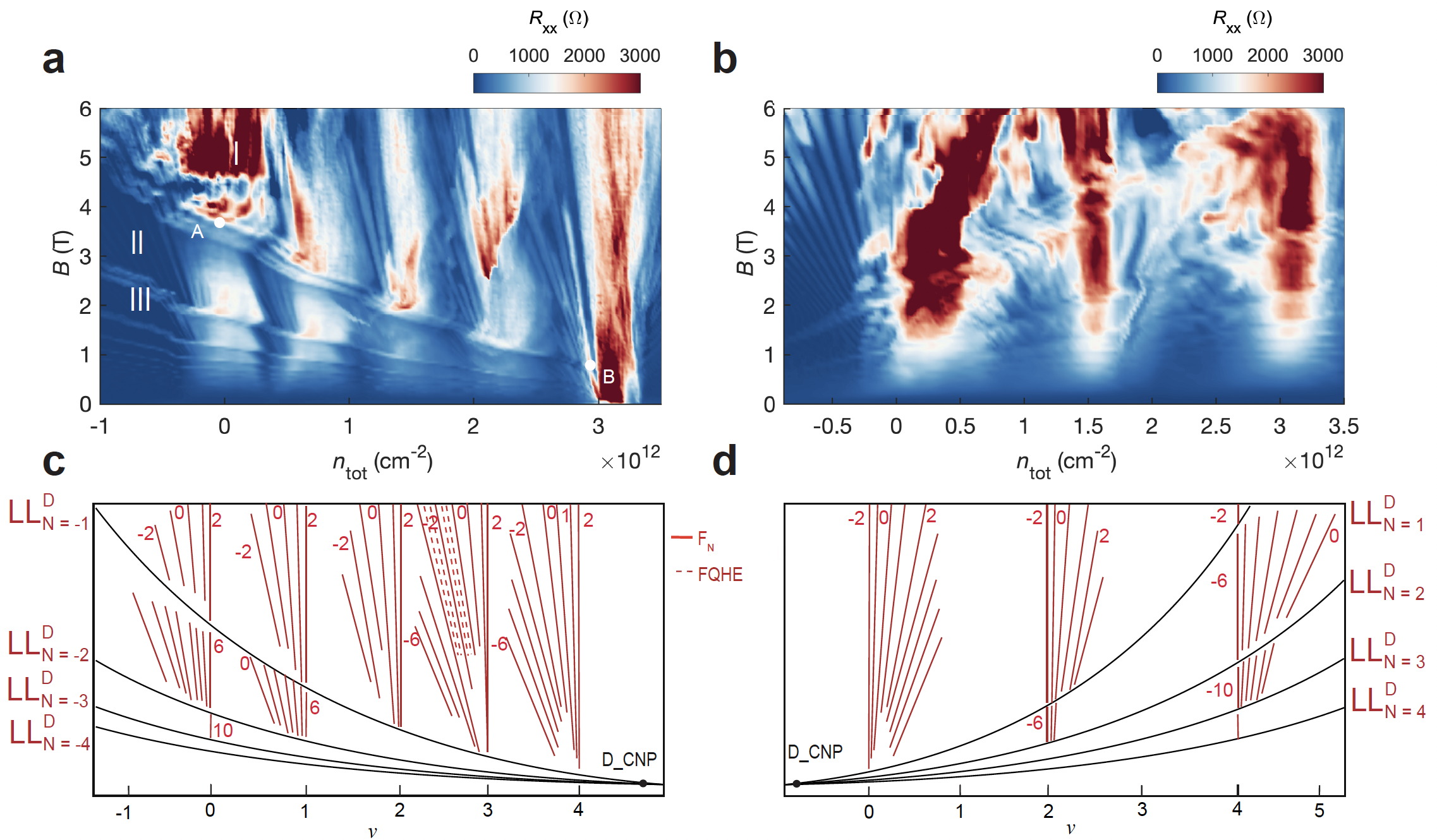}
\caption{$\textbf{Unique Chern number cascade and completely degeneracy lifting of MATMBG flat band.}$ \textbf{a}, Quantum oscillations are shown in the Landau fan diagram for a DMG + MATMBG (1.14° $\pm$ 0.02°) device in a positive displacement field ($D$ = 0.53 V/nm). \textbf{b}, Schematic of the Landau level crossing structure between the flat and Dirac bands. \textbf{c}-\textbf{d}, is the same as a-b, with a reverse displacement field ($D$ = -0.28 V/nm).
}
\label{figS13}
\end{figure*}

\begin{figure*}
\centering
\renewcommand{\thefigure}{S14}
\includegraphics[width= 1\textwidth]{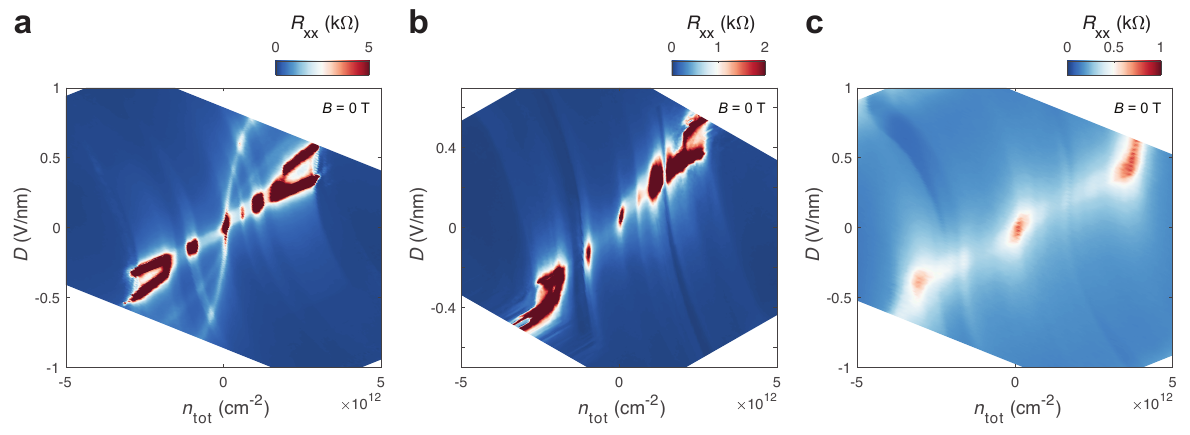}
\caption{$\textbf{Twisted DMG and twisted bilayer graphene (TBG) devices.}$\textbf{a}-\textbf{c}, $n-D$ maps of three DMG stacked on TBG devices (0.94°, 1.01°, 1.20°)  at $B = 0 \, \text{T}$, respectively.
}
\label{figS14}
\end{figure*}


\end{document}